\documentclass[9pt,twocolumn,twoside]{pnas-new}

\templatetype{pnasresearcharticle}

\usepackage{pdfpages}
\usepackage{graphicx}
\usepackage{dcolumn}
\usepackage[normalem]{ulem}
\usepackage{bm}
\usepackage[mathlines]{lineno}
\usepackage{mathtools}
\usepackage{amsmath}
\usepackage{float}
\usepackage[caption=false]{subfig}
\usepackage[english]{babel}
\usepackage{xcolor}
\usepackage{xparse}
\usepackage{hyperref}

\newcommand{\pd}{\partial}
\newcommand{\abs}[1]{\left|#1\right|}
\newcommand{\avg}[1]{\left\langle#1\right\rangle}

\renewcommand{\vec}[1]{\bm{#1}}
\newcommand{\fig}{Fig.~}
\newcommand{\figs}{Figs.~}

\setboolean{displaywatermark}{false}

\newcommand{\ucsb}{Department of Physics, University of California Santa Barbara, Santa Barbara, CA 93106, USA}
\newcommand{\ue}{SUPA, School of Physics and Astronomy, University of Edinburgh, Peter Guthrie Tait Road, Edinburgh EH9 3FD, United Kingdom}
\newcommand{\puc}{Facultad de F{\'i}sica, Pontificia Universidad Cat{\'o}lica de Chile, Santiago 7820436, Chile}

\title{Intercellular Friction and Motility Drive Orientational Order in Cell Monolayers}

\author[a,2]{Michael Chiang}
\author[b]{Austin Hopkins} 
\author[a,c]{Benjamin Loewe}
\author[b,1]{M. Cristina Marchetti}
\author[a,1]{Davide Marenduzzo}

\affil[a]{\ue}
\affil[b]{\ucsb}
\affil[c]{\puc}

\leadauthor{Chiang} 

\significancestatement{Understanding the mechanisms underlying pattern formation and flow in biological tissues is important in embryology and disease. An enigmatic observation in this field is the spontaneous emergence of nematic order in monolayers of nearly-isotropic cells. Here we propose a mechanism for the simultaneous origin of both local nematic and hexatic order. We show that the interplay between cell motility and intercellular friction yields macroscopic spatiotemporal correlations in multicellular flow, which deform and align cells. Our simulations also reveal that cellular geometry implies a fundamental relation between hexatic and nematic defects, which colocalize with cell-cell overlaps and may create hotspots for cellular extrusion.}

\authorcontributions{M.C., A.H., B.L., D.M., and M.C.M. designed research; M.C. and A.H. performed simulations and analyzed data; M.C. and D.M. drafted the paper; M.C., A.H., B.L., D.M., and M.C.M. discussed results and edited the paper.}
\authordeclaration{The authors declare no competing interests.}
\correspondingauthor{\textsuperscript{2}To whom correspondence should be addressed. E-mail: michael.chiang@ed.ac.uk, austinhopkins@ucsb.edu}
\equalauthors{\textsuperscript{1} M.C.M. contributed equally to this work with D.M.}

\keywords{intercellular friction $|$ solid-liquid transitions $|$ nematic and hexatic order $|$ topological defects $|$ cellular extrusion } 

\begin{abstract}
Spatiotemporal patterns in multicellular systems are important to understanding tissue dynamics, for instance, during embryonic development and disease. Here, we use a multiphase field model to study numerically the behavior of a near-confluent monolayer of deformable cells with intercellular friction. Varying friction and cell motility drives a solid-liquid transition, and near the transition boundary, we find the emergence of local nematic order of cell deformation driven by shear-aligning cellular flows. Intercellular friction contributes to the monolayer's viscosity, which significantly increases the spatial correlation in the flow and, concomitantly, the extent of nematic order. We also show that local hexatic and nematic order are tightly coupled and propose a mechanical-geometric model for the colocalization of $+1/2$ nematic defects and $5$-$7$ disclination pairs, which are the structural defects in the hexatic phase. Such topological defects coincide with regions of high cell-cell overlap, suggesting that they may mediate cellular extrusion from the monolayer, as found experimentally. Our results delineate a mechanical basis for the recent observation of nematic and hexatic order in multicellular collectives in experiments and simulations and pinpoint a generic pathway to couple topological and physical effects in these systems.
\end{abstract}

\dates{This manuscript was compiled on \today}
\doi{\url{www.pnas.org/cgi/doi/10.1073/pnas.XXXXXXXXXX}}

\begin{document}

\maketitle
\thispagestyle{firststyle}
\ifthenelse{\boolean{shortarticle}}{\ifthenelse{\boolean{singlecolumn}}{\abscontentformatted}{\abscontent}}{}

\dropcap{T}he collective migration of cells within a biological tissue plays a fundamental role in physiological and pathological processes such as embryogenesis~\cite{Chuai2012}, wound healing~\cite{Poujade2007}, and cancer progression~\cite{Thiery2002,Friedl2009}. A long-standing challenge is to dissect the molecular mechanisms driving such coordinated motion -- from a biophysical standpoint, one aims to understand how the mechanical properties of individual cells and the forces acting upon them give rise to the emergent phenomena seen at the tissue scale~\cite{Ladoux2017}.

Many biological processes, such as tissue development and cancer metastasis, involve a change in cell collective dynamics between a solid-like and a liquid-like state, and have been compared to rigidity and jamming transitions in other soft matter systems~\cite{Atia2021,Lawson-Keister2021}. Modeling has provided fruitful insight into the mechanisms driving these transitions~\cite{LaPorta2019,Alert2020}. For instance, vertex~\cite{Bi2015}, self-propelled Voronoi~\cite{Bi2016,Pasupalak2020}, cellular Potts~\cite{Chiang2016,Durand2019}, and multiphase field models~\cite{Loewe2020,Zhang2020} have successfully captured several defining features of tissue jamming-unjamming, showing how cell shape, deformability and cell-cell adhesion are some of the key determinants of this transition.

Apart from the change in fluidity, another collective phenomenon of tissue monolayers is the spontaneous emergence of orientational nematic order. This has been recently observed in both experiments and simulations~\cite{Saw2017,Kawaguchi2017,Duclos2017,Mueller2019,Zhang2021,Guillamat2022,Zhang2023,Monfared2023,Lin2023}, where nematic order has been measured in terms of the cell shape orientation. Yet, the physical mechanism underlying the origin of cell alignment is still elusive. For example, it is unclear whether contractile or extensile activity is required to create nematic order, nor has it been discussed how robust the order is across the parameter space. The latter aspect is important as, for instance, cells in epithelial tissue in the solid or glass phase are quite isotropic, so any order is by necessity linked with relatively small shape fluctuations. Additionally, as these systems also exhibit bond-orientational (hexatic) order~\cite{Durand2019,Loewe2020,Pasupalak2020,Lin2023}, it is important to understand how nematic and hexatic order interact with each other, and how pervasive the proposed combined ``hexanematic'' order~\cite{Armengol2023,Eckert2023} is in practice.

In this work, we study the dynamics and phase behavior of a confluent monolayer of deformable and motile cells with intercellular and cell-substrate friction. We observe a solid-liquid transition that is accompanied by the emergence of both local hexatic and nematic order. This orientational order is maximal close to the transition and is driven by the onset of a cellular flow, which deforms and aligns cells within a range set by the correlation length of the flow. Moreover, we find that the geometry of our deformable cells creates a coupling between the hexatic and nematic order, such that defects in the hexatic order, corresponding to $5$-$7$ disclination pairs, appear preferentially close to $+1/2$ defects in the nematic order. These defects correlate with regions of enhanced overlap between cells, which provides a mechanism for the topological creation of hotspots for cellular extrusion, as found experimentally~\cite{Saw2017} and in simulations~\cite{Loewe2020,Monfared2023}. Our results therefore provide a mechanical explanation for the emergence of nematic order in cell monolayers, and predict that the latter should be strongly enhanced by intercellular friction. Our finding of an intimate geometric coupling between the hexatic and nematic order provides a mechanical underpinning for the proposal that cellular monolayers can behave, under suitable conditions, as a hexanematic fluid~\cite{Armengol2023,Eckert2023}. They also raise the tantalizing possibility that topology can play an important role in multicellular dynamics and developmental biology.

\section*{Results}

\begin{figure}[!t]
  \centering
  \includegraphics[width=\columnwidth]{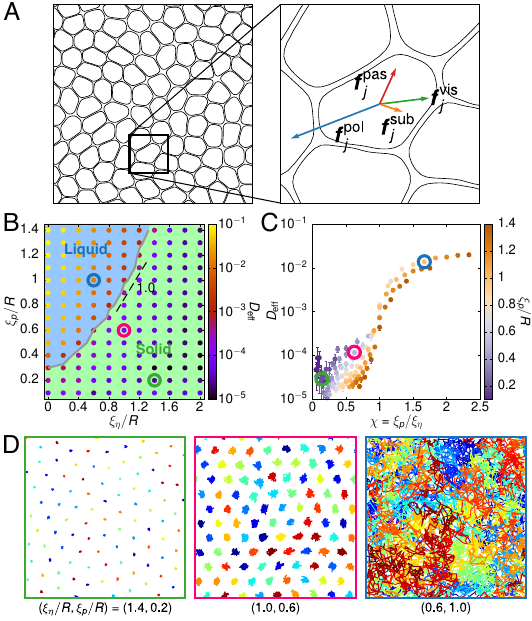}
  \caption{Intercellular friction and motility drive solid-liquid transition in a multiphase field model. (A) A snapshot of the simulated monolayer (\textit{left}) with an enlarged view of one of the cells showing the forces acting upon it (\textit{right}). There are four types of forces: a passive force $\vec{f}_i^{\text{pas}}$ due to the imposed free energy, an active self-propulsion force $\vec{f}_i^{\text{pol}}$, cell-substrate friction $\vec{f}_i^{\text{sub}}$, and cell-cell friction $\vec{f}_i^{\text{vis}}$. (B) Phase diagram showing the effective diffusivity $D_{\text{eff}}$ of the system when varying cell motility $\xi_p$ and the flow screening length $\xi_{\eta}$ (in units of the cell radius $R$). The transition line demarcating the solid and fluid phases is interpolated based on the threshold $D_{\text{eff}} = 10^{-3}$. (C) Points where $\xi_{\eta}/R \geq 0.6$ collapse onto a master curve as a function of the ratio between the persistence and flow screening lengths, $\chi=\xi_p/\xi_{\eta}$. (D) Representative cell trajectories at three points of the phase diagram [corresponding to those circled in (B) and (C)], indicating the transition from solid-like (caging) to fluid-like (neighbor exchange) behavior.}
  \label{fig:model}
\end{figure}

\paragraph*{A Multiphase Field Model for Deformable Cell Monolayers with Cell-Cell and Cell-Substrate Friction.}
We simulate $N$ cells as deformable droplets randomly initialized on a substrate (\fig\ref{fig:model}A). Each cell is modeled by a phase field $\phi_i(\vec{r})$ ($i = 1, ..., N$), where $\phi_i = 1$ marks the cell's interior and $0$ its exterior. In line with previous work~\cite{Mueller2019,Loewe2020,Zhang2020,Hopkins2022,Hopkins2023,Zhang2023}, the droplet's shape is controlled by a Landau-Ginzburg-like free energy (see Materials and Methods), which includes Cahn-Hilliard terms governing the cell's edge tension $\sigma$ and thickness $\xi$, a soft constraint on its area (with radius $R$), and a repulsive term (strength $\epsilon$ with units $[E][L]^{-2}$) that minimizes cell overlaps. Each cell experiences a passive force $\vec{f}_i^{\text{pas}}$ due to the imposed free energy~\cite{Cates2018} and an active self-propulsion force $\vec{f}_i^{\text{pol}}$ (proportional to speed $v_0$) as it undergoes rotational diffusion with rate $D_r$. Importantly, these forces are balanced by cell-cell friction $\vec{f}_i^{\text{vis}}$ (strength $\eta$ with units $[E][T][L]^{-2}$) due to the relative motion between cells and cell-substrate friction $\vec{f}_i^{\text{sub}}$ (strength $\Gamma$ with units $[E][T][L]^{-4}$; see \fig\ref{fig:model}A and Methods for the expressions of these forces). The droplets are evolved over time using advective-relaxational dynamics, and the net effect of the forces on the droplet enters the equations of motion through its advection velocity $\vec{v}_i$.

The overall behavior of the model is tuned by three parameters. First, there is the deformability of individual droplets, $d \equiv \epsilon\xi R/(12\sigma R)$, where at lower $d$ cells remain circular and prefer to overlap rather than deform~\cite{Loewe2020,Hopkins2022,Hopkins2023}. Here, we set $d = 4.16$ such that cells are highly compliant and do not overlap significantly. Second, the cell's motility is tuned by the P\'eclet number $\text{Pe} = \xi_p/R$, where $\xi_p = v_0/D_r$ is the persistence length of cell motion. Third, the relative strength of cell-cell to cell-substrate friction gives a flow screening length $\xi_{\eta} = \sqrt{\eta/\Gamma}$~\cite{Blanch-Mercader2017,Alert2019,Vazquez2022} that quantifies the distance over which the motion of a cell can influence that of another. Throughout this work, we focus on varying $\xi_p$ and $\xi_{\eta}$ to observe their effect on the monolayer dynamics; in particular, we fix cell-substrate friction $\Gamma$ and vary the strength of cell-cell friction $\eta$. We use a system of $N = 100$ and, in selected cases, $400$ cells -- results for the latter are shown in the Supporting Information (SI) unless otherwise stated. We estimate the physical values of the simulation parameters by mapping to data on mammary epithelial MCF-10A cells~\cite{Malinverno2017,Hosseini2020} (see SI), and we find comparable values for the persistence length of motility and velocity correlations between simulations and experiments.

\paragraph{Interplay of Friction and Cell Motility Drives a Solid-Liquid Transition in the Monolayer.}
We first examine the role of cell motility ($\xi_p$) and of the two friction forces ($\xi_{\eta}$) on tissue fluidity. To this end, we measure the mean square displacement (MSD) of individual cells and compute an effective diffusivity $D_{\text{eff}} = \lim_{t\to\infty}\text{MSD}/(4D_0t)$, where $D_0 = v_0^2/(2D_r)$ is the self-diffusivity of an isolated active Brownian particle~\cite{Loewe2020}. \fig\ref{fig:model}B displays a phase diagram of $D_{\text{eff}}$ and shows that, above a critical $\xi_p$, the monolayer undergoes a solid-liquid transition, where cells move from being caged to exchanging neighbors (see \fig\ref{fig:model}D and also \fig{S2} for $N = 400$).

We construct the transition line using the threshold $D_{\text{eff}} = 10^{-3}$, which aligns well with changes in the structural order between the two phases as presented below. The line shows that as $\xi_{\eta}$ increases, or when cell-cell friction dominates over cell-substrate friction, a higher $\xi_p$ is required for the monolayer to melt. Notably, it becomes almost linear when $\xi_{\eta} \ge 0.6$, but is strongly non-linear below this point (\fig\ref{fig:model}B), suggesting there are different mechanisms regulating the melting process. At high $\xi_{\eta}$, the linearity implies that the transition can be described by a single dimensionless parameter $\chi=\xi_p/\xi_{\eta}$, which quantifies the competition of two length scales -- the persistence length of a cell's active propulsion and the correlation length of the motion between cells. Using this definition, points on the phase diagram, at least when $\xi_{\eta} \ge 0.6$, all collapse onto a master curve, and the melting takes place when $\chi\sim 1$ (\fig\ref{fig:model}C and S2). Note that $\chi$ fails to capture the transition accurately when $\xi_{\eta} < 0.6$. Here, the transition line flattens as $\xi_{\eta} \to 0$, suggesting that there is a regime where melting is independent of $\xi_{\eta}$, and other mechanisms, such as cell deformability~\cite{Loewe2020}, may have a larger effect in melting the system.

\begin{figure*}[!t]
  \centering
  \includegraphics[width=\linewidth]{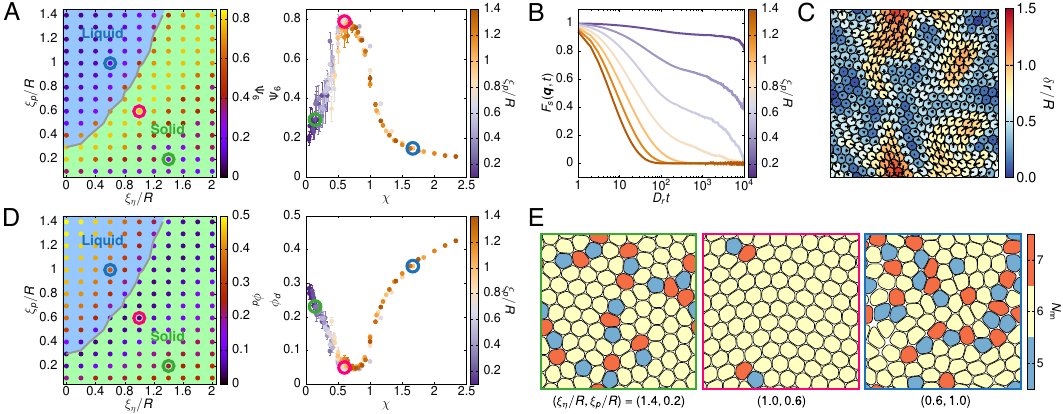}
  \caption{The monolayer exhibits a regime with local hexatic order between a liquid and a glassy, amorphous solid phase. (A) \textit{Left}: Measurements of the global bond-orientational order $\Psi_6$ superposed on the solid-liquid phase diagram shown in \fig\ref{fig:model}B. \textit{Right}: Collapsing $\Psi_6$ based on $\chi$ onto a master curve for points where $\xi_{\eta} \geq 0.6$. (B) Self-intermediate scattering function $F_s(\vec{q},t)$ (with $\abs{\vec{q}} = \pi/R$) when varying $\chi$ (i.e., vary $\xi_p$ at fixed $\xi_{\eta} = 1.0$) for a system of $N = 400$ cells. (C) Cell displacement map of a $400$-cell monolayer at $\xi_{\eta}/R = 1.0$ and $\xi_p/R = 0.6$ over a period of time $D_rt = 100$, corresponding to the timescale when $F_s(\vec{q},t)$ decays to ${\sim}1/2$. Arrows indicate the directions of the cell displacements, with their length twice the magnitude of the actual displacements to aid visualization. (D) Similar to (A), but showing the average fraction of cells with disclinations $\phi_d$ in the monolayer. (E) Representative snapshots of the system at parameter points circled in the phase diagrams in (A) and (D), with cells colored by their number of nearest neighbors $N_{\text{nn}}$, determined by Voronoi tessellation. Here, 5- and 7-fold disclinations are marked in blue and orange, respectively.}
  \label{fig:hexatic}
\end{figure*}

\paragraph{A Local Hexatic Regime Separates the Liquid and Glassy Solid Phases.}
To characterize the structural order of the system, we compute for each cell its bond-orientational order $\psi_{6,j} = \frac{1}{N_{j,\text{nn}}}\sum_{k\in\text{nn}}e^{i6\theta_{jk}}$, where the sum is over the nearest neighbors and $\theta_{jk}$ is the angle between the $x$-axis and the bond vector linking cells $j$ and $k$. The quantity $\Psi_6 = \avg{\abs{\frac{1}{N}\sum_{j=1}^N\psi_{6,j}}}$ then gives the global orientational order of the monolayer. As the system solidifies, $\Psi_6$ first increases to near unity, indicating high order, before it surprisingly decreases to a low value deep in the solid, indicating that the transition is distinct from conventional two-dimensional (2D) melting where the hexatic phase is a liquid (\figs\ref{fig:hexatic}A and S7B). The weakening of $\Psi_6$ deep in the solid regime can be attributed to the random positioning of cells when initializing the monolayer and indicates that the solid regime is a glass, as supported by the behavior of the self-intermediate scattering function and the non-Gaussian parameter, as well as maps of cell displacements (\figs\ref{fig:hexatic}B,C, S4, and S5). These maps show correlated regions that extend over multiple cell widths and are bounded by regions of high strain, as commonly seen in experiments on epithelia. The long relaxational timescale associated with glassy dynamics means that the monolayer remains kinetically frozen in its initial state and is unable to reach a crystalline configuration with lower free energy. Indeed, by starting the simulations with cells arranged on a triangular lattice, $\Psi_6$ approaches unity asymptotically as the system freezes (\fig{S6B}). In line with this, the global translational order $\Psi_T$ of the system (see SI) only changes significantly across the solid-liquid transition when cells are initialized on a lattice (\figs{S3}, S6C, and S7C).

Since the spatial organization of structural defects, i.e., 5- and 7-fold disclinations, plays a prominent role in 2D melting~\cite{Kosterlitz1973,Halperin1978,Nelson1978,Young1979,Durand2019,Loewe2020,Digregorio2022}, we monitor their presence within our model tissue by counting the number of nearest neighbors $N_{\text{nn}}$ of each cell (\figs\ref{fig:hexatic}E, S6A, and S7A). Consistent with $\Psi_6$, the fraction of cells with disclinations $\phi_d$ first decreases as the monolayer freezes but increases again further within the solid regime (\figs\ref{fig:hexatic}D and S7D). We find 5- and 7-fold disclinations are bound in pairs in the intermediate regime of local hexatic order and become unbound in the liquid phase. Some isolated disclinations are seen in the solid phase due to slow, glassy dynamics in relaxing from the initial conditions. While the patterning of defects may appear similar in the glassy solid and liquid phases, the dynamics is different: in the solid regime, $\phi_d$ is static as disclinations are pinned to the same cells, whereas in the liquid, $\phi_d$ fluctuates due to binding-unbinding events (\fig{S8} and Movies S1--S3).

\paragraph{Cellular Flow Promotes Cell Deformation and Local Nematic Order.}
Another topological feature that has attracted lots of interest in tissue sheets is the emergence of nematic order in cell deformation, and the accompanying defects have been implicated in various physiological and pathological processes~\cite{Saw2017,Zhang2021,Guillamat2022,Sarkar2023}. While previous work~\cite{Saw2017,Mueller2019,Zhang2020,Balasubramaniam2021,Zhang2023} has mostly focused on how active stresses, such as individual cell contractility, drive nematic order, here we demonstrate an alternate mechanism by showing that the interplay between intercellular friction and motility can spontaneously give rise to local nematic alignment. In line with recent studies~\cite{Mueller2019,Zhang2020,Hopkins2022,Zhang2023}, we use the shape tensor $\vec{S}_i = -\int d^2\vec{r}\,(\vec{\nabla}\phi_i)(\vec{\nabla}\phi_i)^T$ to determine the cell deformation axis (i.e., the eigenvector corresponding to the largest eigenvalue of $\vec{S}_i$) and define a local nematic order parameter $\Psi_2^L = \avg{\abs{\frac{1}{N}\sum_{j=1}^N\psi_{2,j}}}$, where $\psi_{2,j} = \frac{1}{N_{j,\text{nn}}}\sum_{k\in\text{nn}}e^{i2\theta_{jk}^d}$ and $\theta_{jk}^d$ is the angle between the deformation axes of cells $j$ and $k$. In this way, a higher $\Psi_2^L$ signifies stronger nematic alignment between neighboring cells' deformation axes.

\begin{figure}[!t]
  \centering
  \includegraphics[width=\linewidth]{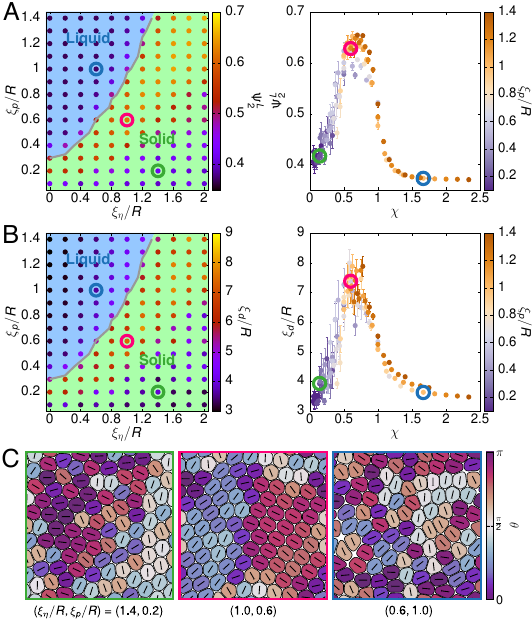}
  \caption{Cell-cell friction drives local nematic alignment of cells. (A) \textit{Left}: Measurements of the local nematic order parameter $\Psi_2^L$ superposed on the phase diagram shown in \fig\ref{fig:model}B. \textit{Right}: Collapsing $\Psi_2^L$ based on $\chi$ onto a master curve for points where $\xi_{\eta} \geq 0.6$. (B) Similar to (A), but for the local characteristic length $\xi_d$ of domains in which cells exhibit coherent nematic alignment. (C) Representative snapshots of the system at points circled in the phase diagrams in (A) and (B), with cells colored by the angle $\theta$ between their deformation axis (the line within each cell) and the $x$-axis.}
  \label{fig:nematic}
\end{figure}

Remarkably, we find the monolayer exhibits an increase in nematic order in the vicinity of the solid-liquid transition, coinciding with the region of the phase space where there is large hexatic order (\fig\ref{fig:nematic}A). This is visually apparent by coloring cells according to their deformation direction, which shows locally aligned domains (\fig\ref{fig:nematic}C and Movies S4--S6). To quantify domain size, we measure the nematic order within a circle of radius $r$, averaged across a set of grid points, and then extract a characteristic length $\xi_d$ from the decay in the order as a function of $r$ (see SI and \fig{S9}). This shows that local alignment can be up to four cell lengths (\fig\ref{fig:nematic}B). These results are largely unaffected by the initial conditions and the system sizes (\figs{S10} and S11).

The emergence of local nematic alignment near the solid-liquid transition can be rationalized by the following argument. First, cellular flow can lead to both cell deformation and cell alignment~\cite{Hernandez2021}. In a coarse-grained model, the rate of change of the deformation tensor should be proportional to the local shear rate, which we can estimate as $\sim v_0/\xi_{\eta}$, with $v_0$ a typical velocity scale. Dimensional analysis then suggests that the flow-induced cell deformation and local nematic order should increase as $\sim v_0/(\xi_{\eta}D_r) = \chi$ (\fig{S12A}). Importantly, this local order can only persist up to the correlation length of the flow, which is proportional to $\xi_{\eta}$ (\fig{S12B}); as a result, the order decreases when the cell persistence length $\xi_p$ becomes larger than $\xi_{\eta}$ (i.e., $\chi > 1$), as cells move away before aligning. These two opposing effects on the local nematic order when increasing $\chi$ therefore argue that the order should be maximal when $\chi \sim 1$, in line with our simulation results.

\begin{figure}[!t]
  \centering
  \includegraphics[width=\linewidth]{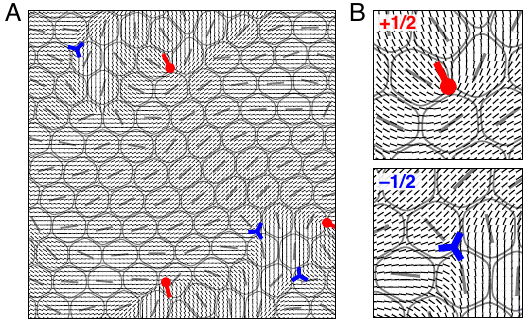}
  \caption{$+1/2$ and $-1/2$ defects emerge near boundaries of local nematic domains. (A) A simulation snapshot showing the deformation axes of individual cells, the coarse-grained director field derived from the $\vec{Q}$ tensor (see SI), and the nematic defects (red tadpoles for $+1/2$ and blue three-edge stars for $-1/2$). (B) An enlarged view of the director field around a $+1/2$ and a $-1/2$ defect.}
  \label{fig:nematic_defect}
\end{figure}

\paragraph{Structural Disclinations Determine the Location of Nematic Defects and Hotspots for Cellular Extrusion.}
We next explore the organization of topological defects in the nematic texture, within the region where $\Psi_2^L$ is large. To this end, we construct a tensor field $\vec{Q}(\vec{r}) = \sum_{i=1}^N \mathcal{W}_i(\vec{r})\,\mathcal{S}_i(2\vec{n}_i\vec{n}_i^T-\vec{I})$, where $\mathcal{S}_i$ is the degree of deformation of cell $i$, $\vec{n}_i$ is its deformation axis, and $\mathcal{W}_i(\vec{r})$ is an ellipsoidal smoothing function (see SI). Nematic defects are then identified by finding local minima in the scalar component of $\vec{Q}$ and computing the topological charge. As shown in \fig\ref{fig:nematic_defect}, $\pm1/2$ defects arise near the boundaries of local nematic domains. 

\begin{figure*}[!t]
  \centering
  \includegraphics[width=\linewidth]{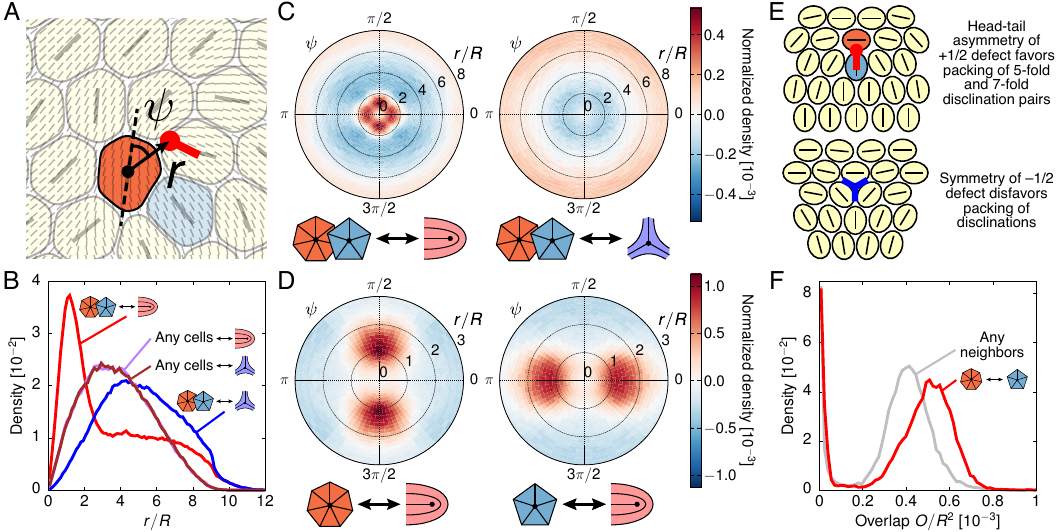}
  \caption{Hexatic $5$-$7$ disclination pairs are strongly correlated with nematic defects. (A) A simulation snapshot illustrating the radial and angular observables ($r$,$\psi$) measured when relating a nematic defect (red tadpole) to its closest hexatic disclination (orange cell). Here, $r$ is the distance between the two defects and $\psi$ is the angle that the cell deformation axis of the hexatic disclination makes with the vector connecting it to the nematic defect core. (B) Probability density functions (PDFs) of the distance $r$ between $\pm1/2$ nematic defects and their closest hexatic disclination ($5$- or $7$-fold), and between $\pm1/2$ defects and a randomly selected cell. (C) Polar heatmaps showing the joint PDFs of $r$ and $\psi$ for the relation between hexatic disclinations and (\textit{left}) $+1/2$ defects or (\textit{right}) $-1/2$ defects. The PDFs are normalized by subtracting the PDF for two randomly selected cells. (D) Similar to (C), but focusing only on the relation between $+1/2$ defects and (\textit{left}) $7$-fold or (\textit{right}) $5$-fold disclinations. (E) Schematics explaining how the geometric layout of the comet-like $+1/2$ nematic defect favors colocalization with $5$-$7$ pairs, whereas the $-1/2$ defect does not. (F) PDFs of the degree of pairwise cell overlap $O_{ij} = \int d^2\vec{r}\,\phi_i^2\phi_j^2$ for a random pair of nearest-neighbor cells (gray curve) and a $5$-$7$ pair (red curve). Results shown here are for the point $(\xi_{\eta}/R,\xi_p/R) = (1.0,0.6)$.}
  \label{fig:disc_defect}
\end{figure*}

Recent studies~\cite{Armengol2023,Eckert2023} have suggested that a tissue monolayer can simultaneously exhibit both hexatic and nematic orientational order, consistent with our results presented above. Here, we aim to establish the physical connection between the topological defects associated with these two types of order -- namely nematic $\pm1/2$ defects and $5$-$7$ disclination pairs, which correspond to dislocations and are the structural defects in the hexatic phase. We first investigate positional correlations between $5$-$7$ pairs and $\pm 1/2$ nematic defect. Surprisingly, we find that $+1/2$ defects are on average significantly closer to $5$-$7$ pairs than $-1/2$ to these pairs (\figs\ref{fig:disc_defect}A,B). The latter are also further away from $5$-$7$ pairs than from a randomly chosen cell within the monolayer.

To gain insight into the mechanisms that drive $+1/2$ defects to form close to $5$-$7$ pairs, we analyze the angular distribution of the relative position of hexatic and nematic defects (\figs\ref{fig:disc_defect}A,C). While $-1/2$ defects are isotropically depleted around $5$-$7$ pairs, the angular distribution of $+1/2$ around $5$-$7$ pairs is anisotropic, with a marked 4-fold symmetry (\fig\ref{fig:disc_defect}C). A separate analysis of 5-fold and 7-fold disclinations shows that they contribute complementary parts of the angular distribution pattern (\fig\ref{fig:disc_defect}D). In particular, the deformation axis of a cell with 7-fold disclination is typically perpendicular to the symmetric axis of a $+1/2$ defect, whereas the deformation axis of a 5-fold disclination is often parallel with it.

Close inspection of simulation snapshots and movies suggests that the difference in the hexatic structure close to $+1/2$ and $-1/2$ defects can be explained geometrically (\fig\ref{fig:disc_defect}E and Movie S7). The head-tail comet asymmetry of the $+1/2$ defect, and its corresponding polar nature, is compatible with cell arrangements that readily accommodate a $5$-$7$ pair. In stark contrast, the $3$-fold symmetry of a $-1/2$ defect is better placed in a region with regular hexagonal packing of cells without structural defects.

This geometric reasoning leads to the expectation that $+1/2$ defects should be associated with larger cell deformations. Previous experimental work~\cite{Saw2017} found that $+1/2$ defects tend to colocalize with regions of high elastic stress and are candidate sites for cellular extrusion from the monolayer. As cell deformation likely correlates with elastic stress, our results suggest a mechanical model for the selection of extrusion hotspots that is driven by the presence of structural $5$-$7$ dislocations, which are attracted by $+1/2$ nematic defects. Accordingly, a quantitative analysis of our multiphase patterns shows that 5-7 dislocations (and hence $+1/2$ defects) are associated with increased cell-cell overlaps (\fig\ref{fig:disc_defect}F), which likely correlate with potential extrusion sites. These results extend previous work showing that extrusion correlates with the location of 5-fold disclinations~\cite{Loewe2020,Monfared2023}, providing a link between topological defects in hexatic and nematic texture.

\section*{Discussions}

In summary, we have used multiphase simulations to study the dynamics and topological structure of a monolayer of motile cells with intercellular friction. When the latter is sufficiently large, a key dimensionless parameter for determining the physical properties of the monolayer is the ratio $\chi$ between the persistence length -- which measures the distance traveled by a cell in isolation before rotational diffusion kicks in -- and the flow screening length -- which measures the correlation length of cellular flow. Increasing $\chi$ triggers a solid-liquid transition~\cite{Bi2015,Chiang2016,Loewe2020}, which can be clearly identified by measuring the effective diffusion coefficients of the cells.

An important outcome of our investigation is that an orientationally ordered phase emerges close to the solid-liquid transition. This phase has both local hexatic and nematic order -- the emergence of the latter is striking given the fact that cells are nearly isotropic in isolation. The appearance of nematic order is due to the onset of correlated cellular flow in the monolayer, as shear forces deform the cells and align them locally~\cite{Hernandez2021}, within a length scale given by the screening length. We note that the tissue viscosity associated with the cell-cell friction employed in our model can be thought of as a proxy for cell-cell adhesion, and recent experiments have indicated that adhesion plays an important role in determining local hexatic and nematic order~\cite{Eckert2023}. The necessity of both cellular flow and a finite flow correlation length explains why nematic order also emerges naturally in monolayers without intercellular friction but with active dipolar forces -- e.g., from the contractile actomyosin cortex. In this case, dipolar activity and motility create flow, while activity and elasticity yield a finite correlation length~\cite{Giomi2015} that determines the size of nematic domains.

Our finding of local orientational order near the solid-liquid transition provides mechanical insight into the observation of a local hexanematic phase in cell monolayers in experiments and simulations~\cite{Armengol2023,Eckert2023}. In particular, we find that geometry leads to an intimate coupling between hexatic and nematic order in the context of topological defects. The shape of cells near comet-like $+1/2$ favors the presence of 5-7 pairs (i.e., structural dislocations), which are the signature of hexatic order; as a result, 5-7 pairs appear near $+1/2$ defects. Previous work~\cite{Nelson1980,Bruinsma1982,Selinger1989} on passive liquid crystals has examined the possible phases in systems with two types of orientational order coupled to each other through suitable terms in a Landau free energy. It would be interesting to ask whether the correlation between nematic defects and 5-7 disclinations reported here can also be understood as the result of an energetic coupling between hexatic and nematic order, and would therefore similarly arise in passive systems.

Finally, the regions with $+1/2$ defects and $5$-$7$ pairs are associated with both increased cell deformation and cell-cell overlap. We therefore suggest that topological patterns provide a mechanism to select these regions as potential sites for cell extrusion from the monolayer, in agreement both with experimental observations that $+1/2$ defects in monolayers are extrusion hotspots~\cite{Saw2017} and with the prediction by simulations that extrusion should occur near 5-fold disclinations in the hexatic order~\cite{Loewe2020,Monfared2023}. We hope our prediction of a geometric coupling between nematic and hexatic defects will stimulate further analysis of the interplay between topology and extrusion in monolayers and model tissues. 

\matmethods{
\paragraph{Model Setup.} We consider a 2D monolayer of $N$ deformable cells, each modeled by a phase field $\phi_i$ ($i = 1, \dots, N$) with periodic boundary conditions. These cells form a monolayer that is nearly confluent, with a packing fraction of ${\sim}0.95$. Similarly to previous work~\cite{Loewe2020,Hopkins2022,Hopkins2023}, the system's total free energy is given by
\begin{align}
\begin{split}
  \mathcal{F} = &\sum_{i=1}^N\bigg[\int d^2\vec{r}\, \kappa\left[\phi_i^2\left(\phi_i-1\right)^2+\xi^2\left(\vec{\nabla}\phi_i\right)^2\right] + \\[-5pt]
  &\lambda A_0\left(1-\int d^2\vec{r}\frac{\phi_i^2}{A_0}\right)^2 + \epsilon\sum_{i<j=1}^N\int d^2\vec{r}\,\phi_i^2\phi_j^2\bigg]\,.
\end{split}
\end{align}
Here, the first two terms fix $\phi_i$ to be close to $1$ within the cell and $0$ otherwise, with $\xi$ controlling the cell's interfacial thickness and $\kappa$ its surface tension $\sigma = \kappa\xi/3$. The third term constrains the cell area to be near $A_0 = \pi R^2$, where $R$ is the ideal cell radius, and the last term enforces steric repulsion between cells by penalizing overlap. Unless otherwise stated, cells are initially positioned randomly as circular droplets with radius $R$, and the phase fields are evolved over time using advective-relaxational dynamics
\begin{align}
  \pd_t\phi_i + \vec{v}_i\cdot\vec{\nabla}\phi_i = -\mu_i\,, \label{eqn:model_A}
\end{align}
where $\mu_i = \delta\mathcal{F}/\delta\phi_i$ is a chemical potential and $\vec{v}_i$ the advection velocity of a cell.

We assume the dynamics to be overdamped and determine $\vec{v}_i$ through a force balance equation that includes four types of forces:
\begin{align}
  \vec{f}_i^{\text{pas}} + \vec{f}_i^{\text{pol}} + \vec{f}_i^{\text{sub}} + \vec{f}_i^{\text{vis}} = \vec{0}. \label{eqn:force_balance}
\end{align}
Specifically, there is a passive force 
\begin{align}
  \vec{f}_i^{\text{pas}} = -\sum_{j=1}^N\int d^2\vec{r}\,\phi_i\phi_j\vec{\nabla}\mu_j 
\end{align}
as a result of the imposed free energy $\mathcal{F}$. There is a self-propulsion force that accounts for the cell's motility,
\begin{align}
  \vec{f}_i^{\text{pol}} = \Gamma v_0\int d^2\vec{r}\,\phi_i\vec{P}(\vec{x})\,,
\end{align}
where $\Gamma$ is a damping constant due to the substrate and $\vec{P}(\vec{x}) = \Phi^{-1}\sum_{j=1}^N\phi_j\vec{p}_j$ is the tissue polarization~\cite{Chiang2024}. Here, $\Phi = \sum_{k=1}^N\phi_k$ and $\vec{p}_j = (\cos\theta_j,\sin\theta_j)$ is the cell's propulsion direction, which is assumed to undergo rotational diffusion with rate $D_r$ [i.e., $d\theta_j = \sqrt{2 D_r}\,dW_j(t)$, where $W_j$ is a Wiener process]. Finally, there are two types of friction force: one between the monolayer and the substrate, characterized by 
\begin{align}
  \vec{f}_i^{\text{sub}} = -\Gamma\int d^2\vec{r}\,\phi_i\vec{V}(\vec{x})\,,
\end{align}
with $\vec{V}(\vec{x}) = \Phi^{-1}\sum_{j=1}^N\phi_j\vec{v}_j$ the tissue velocity field~\cite{Chiang2024}, and another one between the cells, expressed as
\begin{align}
  \vec{f}_i^{\text{vis}} = \eta\sum_{j=1}^N\int d^2\vec{r}\,\vec{\mathcal{I}}(\phi_i,\phi_j) \cdot \left[\vec{V}(\vec{x})-\vec{v}_j\right]\,,
\end{align}
where 
\begin{align}
  \mathcal{I}_{\alpha\beta}(\phi_i,\phi_j) = \frac{1}{\Phi}\left[(\pd_{\gamma}\phi_i)(\pd_{\gamma}\phi_j)\delta_{\alpha\beta}+(\pd_{\alpha}\phi_i)(\pd_{\beta}\phi_j)\right]
\end{align}
is a tensor related to the degree of interfacial overlap between cells $i$ and $j$ (i.e., cells only experience friction when they are close to each other). This expression for the cell-cell friction can be derived by considering the tissue as a viscous, compressible medium that can swell with fluid intake (hence the appearance of both bulk and shear terms in $\vec{\mathcal{I}}$)~\cite{Chiang2024}. One can show that, in the limit where three-field overlaps are rare, this form of friction can be approximated as a sum of pairwise friction proportional to $\vec{v}_i - \vec{v}_j$. To gain some intuition of this friction force, in \fig{S1} we show how varying it affects $\vec{v}_i$ during the head-on collision between two cells. A full list of the parameter values used in this work and the numerical procedure for solving for $\vec{v}_i$ and $\phi_i$ are provided in the SI.

\paragraph{Data, Materials, and Software Availability.} Simulation code and data for the figures have been deposited in Edinburgh DataShare (\url{https://doi.org/10.7488/ds/7799}).
}

\showmatmethods{}

\acknow{A.H. and M.C.M. were supported by the National Science Foundation Grant No. DMR-2041459. This research has received funding (B. L.) from the European Research Council under the European Union’s Horizon 2020 research and innovation programme (Grant Agreement No. 851196).}

\showacknow{}

\bibliography{viscous_phase_field}

\clearpage



\section*{Supplementary material}

# Intercellular Friction and Motility Drive Orientational Order in Cell Monolayers: Supplemental Materials

Michael Chiang,[1, *] Austin Hopkins,[2] Benjamin Loewe,[1] M. Cristina Marchetti,[2, †] and Davide Marenduzzo[1, †]

[1]*SUPA, School of Physics and Astronomy, University of Edinburgh, Peter Guthrie Tait Road, Edinburgh EH9 3FD, United Kingdom*

[2]*Department of Physics, University of California Santa Barbara, Santa Barbara, CA 93106, USA*

(Dated: November 1, 2023)

## SUPPLEMENTAL INFORMATION ON SIMULATION METHODS

In the following, we first discuss the derivation of the chemical potential and the advection velocity of each cell, as they are involved in the equations of motion of the phase fields. We then provide additional information on the simulation procedure, as well as the full list of model parameters and the values used. Finally, we explain more in detail the definition and computation of some of the structural and dynamical observables measured in the simulations.

### Computing the Chemical Potential $\mu_i$

Within our monolayer model, it is necessary to explicitly compute the chemical potential of a cell $\mu_i = \delta\mathcal{F}/\delta\phi_i$, as it is required for updating the passive force $\boldsymbol{f}_i^{\mathrm{pas}}$ acting upon the cell and for evolving the phase field $\phi_i$ over time. Recall that the free energy of the system is given by

$$\mathcal{F} = \sum_{i=1}^{N} \left\{ \int d^2\boldsymbol{r}\, \kappa \left[ \phi_i^2(\phi_i-1)^2 + \xi^2(\boldsymbol{\nabla}\phi_i)^2 \right] + \lambda A_0\, \delta A_i[\phi_i] + \epsilon \sum_{i<j=1}^{N} \int d^2\boldsymbol{r}\, \phi_i^2\phi_j^2 \right\}, \tag{S1}$$

with $A_0 = \pi R^2$ and

$$\delta A_i[\phi_i] = 1 - \int d^2\boldsymbol{r}\, \frac{\phi_i^2}{A_0}. \tag{S2}$$

Taking the functional derivative with respect to $\phi_i$, we find

$$\mu_i = \frac{\delta\mathcal{F}}{\delta\phi_i} = 2\kappa \left[ \phi_i(\phi_i-1)(2\phi_i-1) - \xi^2(\nabla^2\phi_i) \right] - 4\lambda\phi_i\delta A_i + 2\epsilon\phi_i \left[ h(\boldsymbol{x}) - \phi_i^2 \right], \tag{S3}$$

where $h(\boldsymbol{x}) = \sum_{j=1}^{N} \phi_j^2(\boldsymbol{x})$ is an auxiliary field introduced to enable parallel computation of individual phase fields.

### Solving for the Cell Advection Velocity $\boldsymbol{v}_i$

As discussed in the main text, we impose force balance within the model [Eq. (3)], which can be written as

$$-f_{i,\alpha}^{\mathrm{sub}}(\{\phi_j\}, \{\boldsymbol{v}_j\}) - f_{i,\alpha}^{\mathrm{vis}}(\{\phi_j\}, \{\boldsymbol{v}_j\}) = f_{i,\alpha}^{\mathrm{pas}}(\{\phi_j\}) + f_{i,\alpha}^{\mathrm{pol}}(\{\phi_j\}). \tag{S4}$$

Here, we use Einstein summation for the Greek indices denoting Cartesian components, whereas there is no implicit summation for the Latin cell indices. To solve for the advection velocity, we write all velocity-dependent forces explicitly in terms of $v_{i,\alpha}$. The challenge in doing so lies in the intercellular friction term, which contains the total tissue velocity field $V_\alpha$, i.e.,

$$f_{i,\alpha}^{\mathrm{vis}} = -\eta \sum_{j=1}^{N} \int d^2\boldsymbol{r}\, \frac{1}{\Phi} \left[ (\partial_\gamma\phi_i)(\partial_\gamma\phi_j)\delta_{\alpha\beta} + (\partial_\alpha\phi_i)(\partial_\beta\phi_j) \right] (v_{j,\beta} - V_\beta). \tag{S5}$$

To make progress, we observe that

$$\sum_{j=1}^{N} (\partial_\beta \phi_j) \, V_\beta = \sum_{j=1}^{N} (\partial_\beta \phi_j) \sum_{k=1}^{N} \frac{\phi_k}{\Phi} \, v_{k,\beta} = \sum_{j=1}^{N} \frac{\phi_j}{\Phi} (\partial_\beta \Phi) \, v_{j,\beta} \,, \tag{S6}$$

where we have swapped the summing indices in the second equality. Hence, one can write

$$\sum_{j=1}^{N} \int d^2 \boldsymbol{r} \, \frac{\partial_\alpha \phi_i}{\Phi} (\partial_\beta \phi_j)(v_{j,\beta} - V_\beta) = \sum_{j=1}^{N} \int d^2 \boldsymbol{r} \, \frac{\partial_\alpha \phi_i}{\Phi} \left[ (\partial_\beta \phi_j) - \frac{\phi_j (\partial_\beta \Phi)}{\Phi} \right] v_{j,\beta} \equiv \sum_{j=1}^{N} K_{ij,\alpha\beta} \, v_{j,\beta} \,, \tag{S7}$$

and, by the same procedure,

$$\sum_{j=1}^{N} \int d^2 \boldsymbol{r} \, \frac{\partial_\gamma \phi_i}{\Phi} (\partial_\gamma \phi_j) \delta_{\alpha\beta}(v_{j,\beta} - V_\beta) = \sum_{j=1}^{N} K_{ij,\gamma\gamma} \delta_{\alpha\beta} \, v_{j,\beta} \,, \tag{S8}$$

such that

$$f_{i,\alpha}^{\text{vis}} = -\eta \sum_{j=1}^{N} \left( K_{ij,\gamma\gamma} \delta_{\alpha\beta} + K_{ij,\alpha\beta} \right) v_{j,\beta} \,. \tag{S9}$$

Noting that the cell-substrate friction term can be written as

$$f_{i,\alpha}^{\text{sub}} = -\Gamma \sum_{j=1}^{N} \int d^2 \boldsymbol{r} \, \frac{\phi_i \phi_j}{\Phi} \, v_{j,\alpha} = -\Gamma \sum_{j=1}^{N} \hat{O}_{ij} \delta_{\alpha\beta} \, v_{j,\beta} \,, \tag{S10}$$

with

$$\hat{O}_{ij} \equiv \int d^2 \boldsymbol{r} \, \frac{\phi_i \phi_j}{\Phi} \,, \tag{S11}$$

the force balance equation can therefore be expressed as

$$\sum_{j=1}^{N} \left[ (\eta K_{ij,\gamma\gamma} + \Gamma \hat{O}_{ij}) \delta_{\alpha\beta} + \eta K_{ij,\alpha\beta} \right] v_{j,\beta} = f_{i,\alpha}^{\text{pas}} + f_{i,\alpha}^{\text{pol}} \equiv u_{i,\alpha} \,, \tag{S12}$$

where

$$u_{i,\alpha} = -\sum_{j=1}^{N} \int d^2 \boldsymbol{r} \, \phi_i \phi_j (\partial_\alpha \mu_j) + \Gamma v_0 \sum_{j=1}^{N} \hat{O}_{ij} \, p_{j,\alpha} \,. \tag{S13}$$

This is a set of linear equations in $v_{j,\beta}$ that can be cast in a matrix form. Specifically, by defining

$$\boldsymbol{A}_{ij} \equiv \left[ \eta \, \text{Tr}(\boldsymbol{K}_{ij}) + \Gamma \hat{O}_{ij} \right] \boldsymbol{I} + \eta \, \boldsymbol{K}_{ij} \tag{S14}$$

and writing $\boldsymbol{v}_i = (v_{i,x}, v_{i,y})^T$ and $\boldsymbol{u}_i = (u_{i,x}, u_{i,y})^T$, Eq. (S12) can be written as

$$\begin{pmatrix} \boldsymbol{A}_{11} & \boldsymbol{A}_{12} & \cdots & \boldsymbol{A}_{1N} \\ \boldsymbol{A}_{21} & \boldsymbol{A}_{22} & \cdots & \boldsymbol{A}_{2N} \\ \vdots & \vdots & \ddots & \vdots \\ \boldsymbol{A}_{N1} & \boldsymbol{A}_{N2} & \cdots & \boldsymbol{A}_{NN} \end{pmatrix} \begin{pmatrix} \boldsymbol{v}_1 \\ \boldsymbol{v}_2 \\ \vdots \\ \boldsymbol{v}_N \end{pmatrix} = \begin{pmatrix} \boldsymbol{u}_1 \\ \boldsymbol{u}_2 \\ \vdots \\ \boldsymbol{u}_N \end{pmatrix} \,, \tag{S15}$$

which can be inverted numerically to solve for $v_{i,\alpha}$.

## Additional Details of the Simulation Procedure

We consider different initial conditions and system sizes for the simulations. For $N = 100$ cells, we examine both the case where cells are randomly initialized (with box size $L_x = L_y = 145$) and the case where they are initially positioned on a triangular lattice (spacing $2R$, and the box dimensions are $L_x = 156$ and $L_y = 135$; this aspect ratio preserves the regularity of the lattice). For $N = 400$, we only perform simulations for a random initial configuration (with size $L_x = L_y = 290$; see Fig. S2 for snapshots of these different initial conditions and system sizes). The box dimensions are chosen to give a pack fraction of approximately 0.95. The results presented in the main text are for the case of $N = 100$ cells that are initialized randomly, whereas the results for other conditions are shown here in the SM.

We solve the matrix equations for $\boldsymbol{v}_i$ [Eq. (S15)] using standard `LAPACK` routines, and we numerically integrate the equations of motion for $\phi_i$ [Eq. (2)] using a third-order upwind finite difference scheme. To increase computational efficiency, we employ the domain decomposition strategy when updating individual phase fields, and the simulations are parallelized using `OpenMP`. The full details of this parallelization strategy have been discussed in our previous work [15].

In each simulation, the system is first allowed to relax and equilibrate for $10^5$ timesteps, during which cell motility and friction forces are switched off. Afterwards, all forces are turned on, and the simulation runs for $10^6$ timesteps for further equilibration. Finally, the simulation is run for $10^7$ timesteps with the structural and dynamical properties of the monolayer sampled every $10^3$ timesteps. We perform multiple runs (10 for $N = 100$ and 5 for $N = 400$) for each parameter point of the phase diagram. Unless otherwise stated, ensemble averages $\langle \cdot \rangle$ of observables are computed by averaging over both time and simulation runs.

## Simulation Parameters

The following table lists the full set of model parameters and the values used in the simulations.

| Parameter | Interpretation | Dimensions | Value(s) |
|:---:|:---:|:---:|:---:|
| $\kappa$ | Surface tension | $[E][L]^{-2}$ | 0.006 |
| $\xi$ | Interfacial thickness | $[L]$ | 1 |
| $\lambda$ | Area constraint | $[E][L]^{-2}$ | 0.1 |
| $\epsilon$ | Cell-cell repulsion | $[E][L]^{-2}$ | 0.1 |
| $R$ | Ideal cell size | $[L]$ | 8 |
| $D_r$ | Rotational diffusion rate | $[T]^{-1}$ | 0.001 |
| $\Gamma$ | Cell-substrate friction | $[E][T][L]^{-4}$ | 0.0156 |
| $\eta$ | Cell-cell friction | $[E][T][L]^{-2}$ | 0.0–4.0 |
| $v_0$ | Cell motility | $[L][T]^{-1}$ | 0.0008–0.0112 |
| $\delta x$ | Size of each lattice pixel | $[L]$ | 1 |
| $\delta t$ | Size of each timestep | $[T]$ | 1 |

Table S1. A full list of the model parameters, their dimensions, and the values used in the simulations. Note that length is expressed in units of the grid-point spacing $\delta x$, time is measured in units of the simulation timestep $\delta t$, and energy is measured in units of $\delta E$.

## Characterizing the Monolayer's Structure and Dynamics in the Solid-Liquid Transition

When analyzing the solid-liquid transition, we classify the monolayer to be in the solid or liquid phase based on the average effective diffusivity $D_{\mathrm{eff}}$ of the cells. In practice, this is computed by performing linear fits to the mean square displacement curves at late times ($D_r t = 4000$ to $8000$) and using the slope of the fit to estimate the diffusivity. We then use the threshold $D_{\mathrm{eff}}^* = 10^{-3}$ to distinguish between the solid and the liquid phase, as this matches well with results from structural observables.

As part of the analysis of the monolayer's structure, we compute the translational order parameter $\Psi_T$, in addition to the bond-orientational (hexatic) order parameter $\Psi_6$. This parameter is defined as

$$\Psi_T = \left\langle \left| \frac{1}{N} \sum_{j=1}^{N} \exp\left(i\boldsymbol{q} \cdot \boldsymbol{r}_j\right) \right| \right\rangle, \tag{S16}$$

where $\boldsymbol{r}_j$ is the center of mass of cell $j$ and $\boldsymbol{q}$ is a reciprocal lattice vector. The ensemble average here is computed over the reciprocal lattice vectors for a regular triangular lattice, i.e., $\boldsymbol{q}_1 = \frac{\pi}{R}\left(1, \frac{1}{\sqrt{3}}\right)$ and $\boldsymbol{q}_2 = \frac{\pi}{R}\left(0, \frac{2}{\sqrt{3}}\right)$, as well as over time and simulation runs. Note that $\Psi_T \sim 1$ when the system exhibits crystalline ordering, and this is only found in our simulations when cells within the monolayer are initialized on a lattice (see Figs. S3, S4B, and S5B).

To quantify glassy dynamics, we compute two observables. One of them is the self-intermediate scattering function $F_s(\boldsymbol{q}, t)$, which monitors the relaxation of the system's structural ordering and is written as

$$F_s(\boldsymbol{q}, t) = \left\langle \left| \frac{1}{N} \sum_{j=1}^{N} \exp\left[i\boldsymbol{q} \cdot \left[\boldsymbol{r}_j(t) - \boldsymbol{r}_j(0)\right]\right] \right| \right\rangle, \tag{S17}$$

where the average is taken over multiple wavevectors $\boldsymbol{q}$ with the same magnitude $|\boldsymbol{q}| = \pi/R$ and over lag time and simulation runs. Systems with glassy dynamics typically exhibit two relaxation regimes, as observed in our simulations (see Fig. 2B). The other observable is the non-Gaussian parameter $\alpha_2$ and is given by

$$\alpha_2 = \frac{1}{2} \frac{\left\langle \frac{1}{N} \sum_{i=1}^{N} \Delta \boldsymbol{r}_i^4 \right\rangle}{\left\langle \frac{1}{N} \sum_{i=1}^{N} \Delta \boldsymbol{r}_i^2 \right\rangle^2} - 1, \tag{S18}$$

where $\Delta \boldsymbol{r}_i = \boldsymbol{r}_i(t) - \boldsymbol{r}_i(0)$ and the average is taken over lag time and simulation runs. This typically peaks in the parameter regime where cells are experiencing transient caging effects.

### Quantifying the Size of Local Nematic Domains in Cell Deformation and Cellular Flow

To estimate the size of local domains where we find nematic alignment in the cell, we first define a set of rectangular grid points $G$ within the monolayer ($10 \times 10$ points, equally spaced; note that these points generally do not coincide with cell positions). For each grid point $p$, we then compute the degree of nematic ordering of the deformation axes $\Psi_d$ and of the advection velocities $\Psi_v$ of all the cells within a threshold radius $r$ from the point (we use the distance between each cell's center of mass and the grid point; see Fig. S7A), and we average this result across all grid points. Mathematically, this local order can be expressed as

$$\Psi_{d,v}(r) = \left\langle \frac{1}{N_p} \sum_{p \in G} \left| \frac{1}{N_r^p} \sum_{j=1}^{N} \Theta(r - |\boldsymbol{r}_j - \boldsymbol{r}_p|) \exp\left(i2\theta_j^{d,v}\right) \right| \right\rangle, \tag{S19}$$

where $\theta_j^{d,v}$ is the angle that the deformation axis or the advection velocity of cell $j$ makes with the $x$-axis, $\boldsymbol{r}_p$ is the position of the grid point $p$, $N_p$ is the number of grid points, $N_r^p = \sum_{k=1}^{N} \Theta(r - |\boldsymbol{r}_k - \boldsymbol{r}_p|)$, and $\Theta(x)$ is the Heaviside step function. To determine the domain size, we vary $r$ to obtain a decay curve of $\Psi_{d,v}$ as a function of this threshold (Figs. S7B,D). Finally, we define a threshold in the ordering (based on the maximum global order observed when considering the entire system) to obtain a characteristic length $\xi_{d,v}$ (Figs. S7C,E).

### $\boldsymbol{Q}$ Tensor Definition, the Director Field, and Nematic Defects Identification

As mentioned in the results section, the $\boldsymbol{Q}$ tensor is defined as

$$Q_{\alpha\beta}(\boldsymbol{r}) = \sum_{i=1}^{N} \mathcal{S}_i(\boldsymbol{r}) \mathcal{D}_i(2n_{i,\alpha} n_{i,\beta} - \delta_{\alpha\beta}). \tag{S20}$$

Here, $\mathcal{D}_i$ is the degree of deformation of cell $i$ and is given by

$$\mathcal{D}_i = \frac{\lambda_{i,+} - \lambda_{i,-}}{\lambda_{i,+} + \lambda_{i,-}} \, , \tag{S21}$$

where $\lambda_{i,\pm}$ are the eigenvalues of the shape tensor $\boldsymbol{S}_i$, and $\boldsymbol{n}_i$ is the eigenvector associated with $\lambda_{i,+}$ (the deformation axis). $\mathcal{S}_i(\boldsymbol{r})$ is an ellipsoidal smoothing function defined as

$$\mathcal{S}_i(\boldsymbol{r}) = \frac{1}{2} \left[ 1 + \tanh \left[ 1 - \left( \frac{x'}{\lambda_{i,+}} \right)^2 - \left( \frac{y'}{\lambda_{i,-}} \right)^2 \right] \right] , \tag{S22}$$

where $x'$ and $y'$ are Cartesian coordinates in the frame where the $x$ axis aligns with the deformation axis of the cell. When plotting the director field and identifying nematic defects, we further smooth the $\boldsymbol{Q}$ field using a $3R \times 3R$ rolling window – i.e., we center the window at each grid point and replace $Q_{\alpha\beta}$ at that point by the average over all points within the window. To locate defects, we first mark the positions of the local minima in the eigenvalue field derived from this smoothed $\boldsymbol{Q}$ as candidate defects, since these will be the locations where the tissue is close to an isotropic state. Next, the topological charge is computed around each candidate defect using a square contour of width $3\,\delta x$, and only those that give a charge consistent with that of a nematic defect are retained. To reduce spurious detection, candidates (with the same topological charge) that are closer than a threshold of $2.5\,\delta x$ are clustered and treated as a single defect. This defect identification algorithm is highly robust, as the resulting set of defects obeys the conservation of topological charge (which is zero for a 2D periodic planar surface).

## SUPPLEMENTAL FIGURES

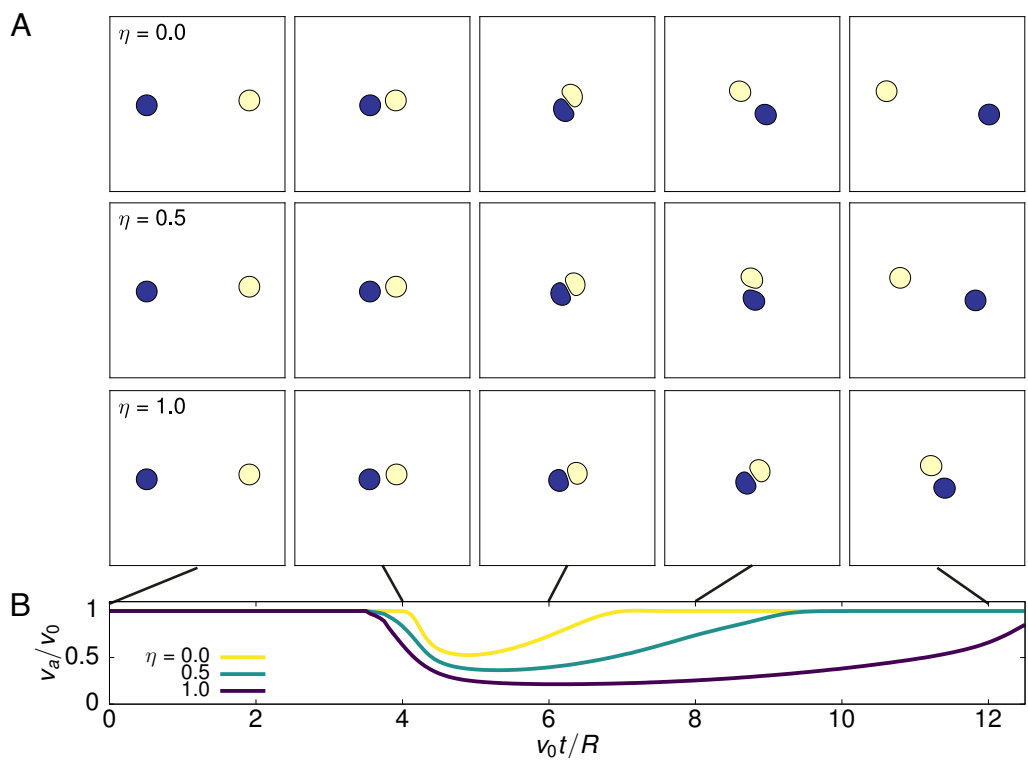

Figure S1. Two-body collision simulations highlighting the effect of cell-cell friction (strength $\eta$). (A). Simulation snapshots at several time points during the collision process, for three $\eta$ values. Here, the two cells move towards each other with a self-propulsion force that is proportional to $v_0 = 0.005$. The cells are initialized with their center of mass slightly off-centered in the $y$ direction (by two grid points), so that they can still move past each other when colliding. (B) The magnitude of the advection velocity $v_a$, normalized by $v_0$, of the cells during the collision (a similar result is observed for the center of mass velocity). Note that the cells slow down more and the collision duration increases as the friction strength $\eta$ becomes stronger.

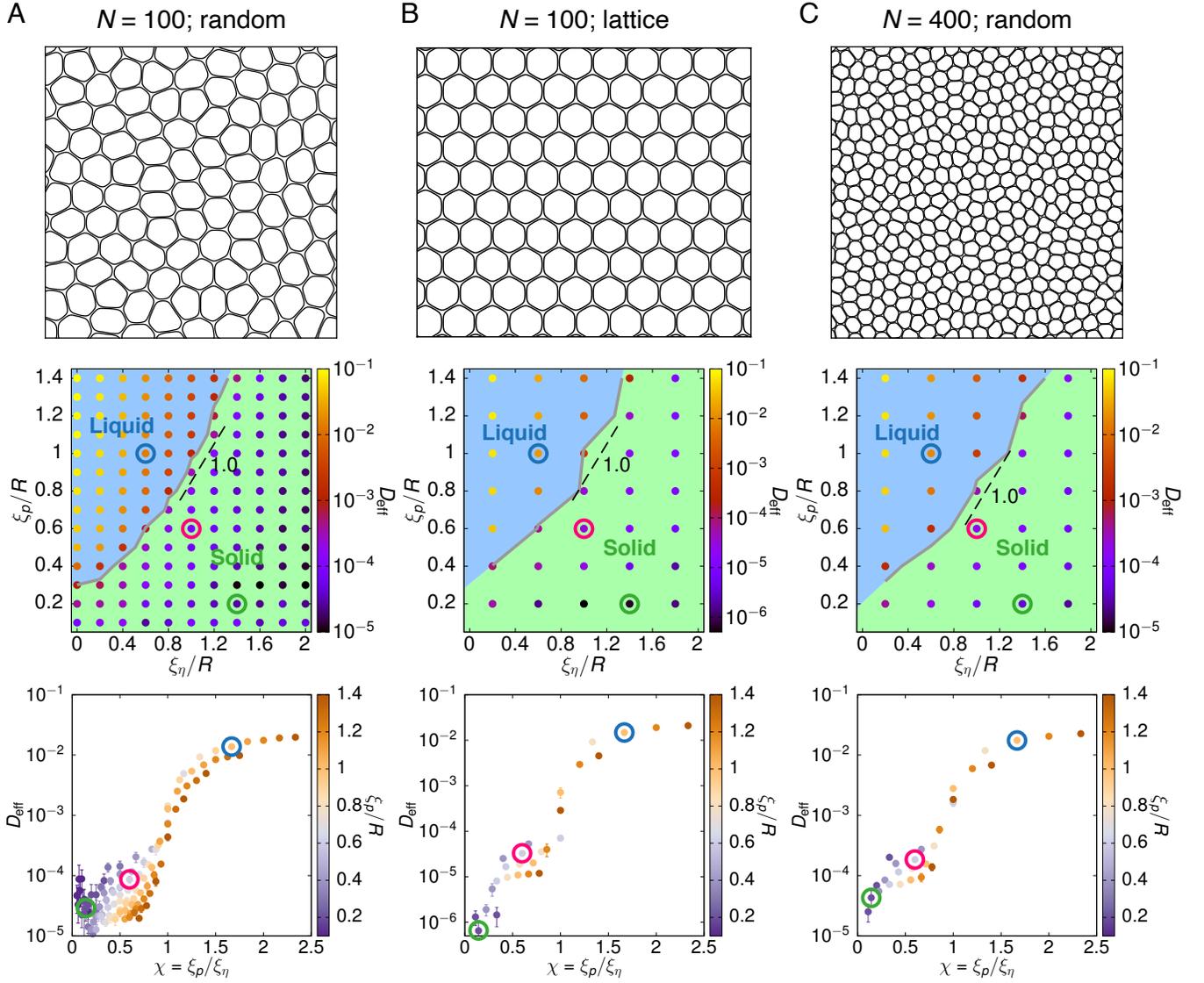

Figure S2. Comparing the solid-liquid phase diagram for three different system setups. (A) $N = 100$ cells that are initialized randomly. (B) $N = 100$ cells that are initialized on a triangular lattice. (C) $N = 400$ cells that are initialized randomly. *Top*: the initial configuration of the cells. *Middle*: the phase diagram, with points colored by the measured effective diffusivity $D_{\text{eff}}$, and the transition boundary is based on the threshold $D_{\text{eff}} = 10^{-3}$. *Bottom*: collapsing $D_{\text{eff}}$ based on $\chi = \xi_p/\xi_\eta$ onto a master curve, for points where $\xi_\eta \geq 0.6$.

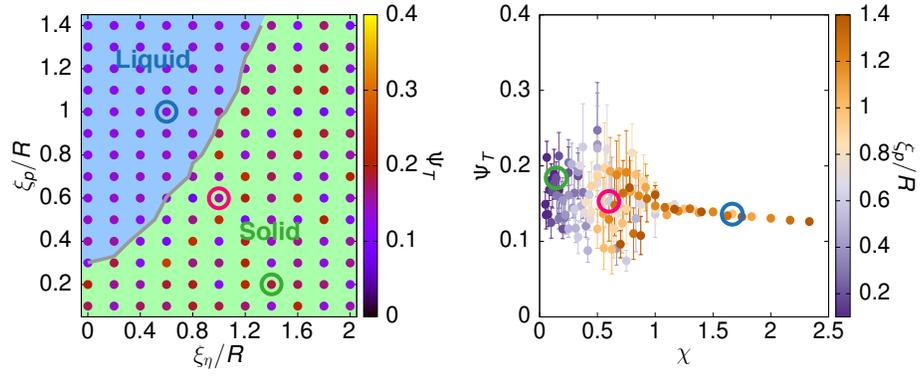

Figure S3. Global translational order $\Psi_T$ of the system ($N = 100$ cells, randomly initialized). *Left*: Measurements of $\Psi_T$ superposed on the solid-liquid phase diagram shown in Fig. 1B (and Fig. S2A). *Right*: Collapsing $\Psi_T$ based on $\chi$ onto a master curve, for points where $\xi_\eta \geq 0.6$.

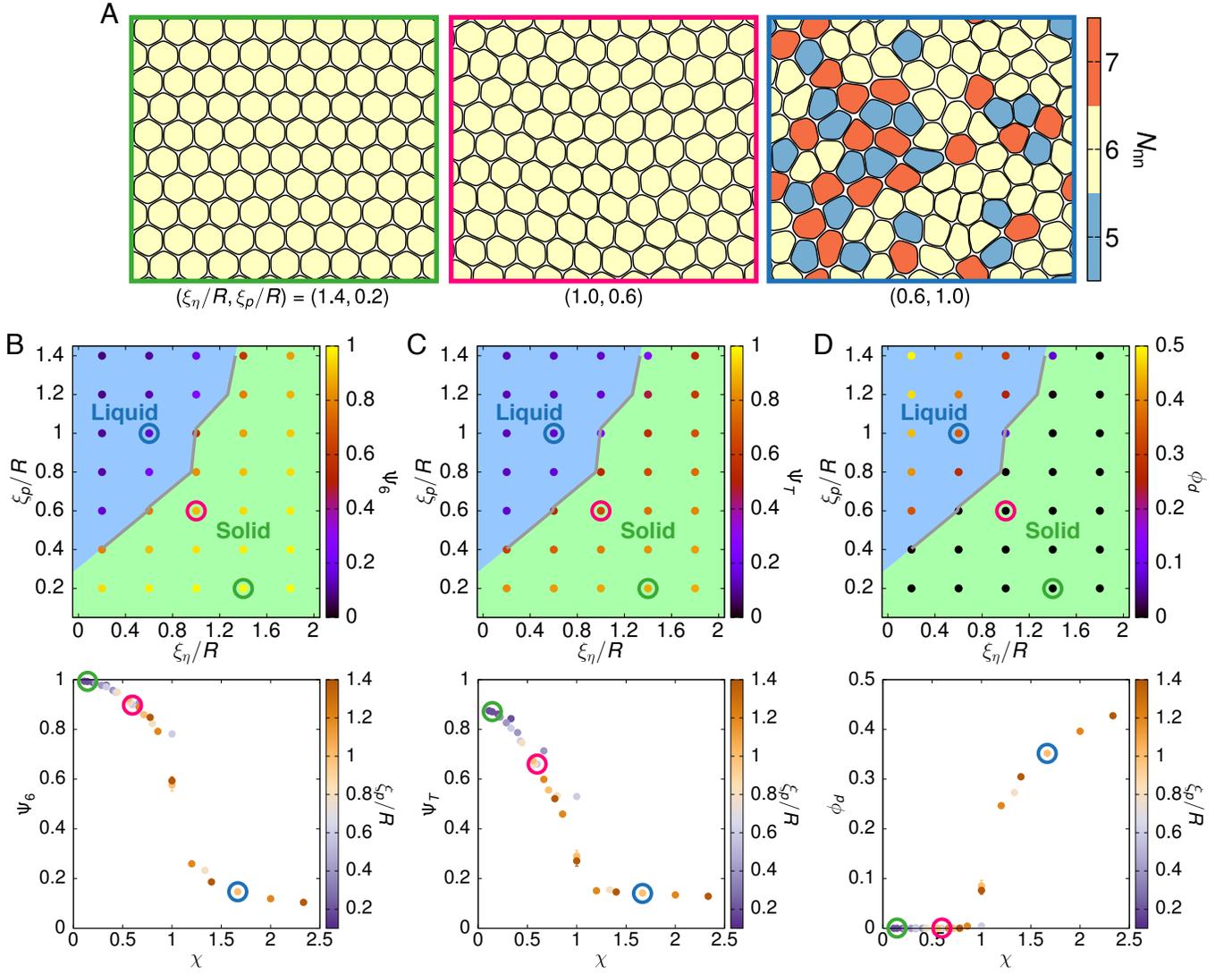

Figure S4. Quantifying the degree of hexatic and translational order for the case of $N = 100$ cells that are initialized on a triangular lattice. (A) Simulation snapshots of three points in the phase space, with cells colored by their number of nearest neighbors $N_{nn}$ (i.e., 5- and 7-fold disclinations are marked in blue and orange, respectively). (B)–(D) *Top*: Measurements of the (B) hexatic order parameter $\Psi_6$, (C) translational order parameter $\Psi_T$, and (D) fraction of cells with disclinations $\phi_d$ superposed on the solid-liquid phase diagram shown in Fig. S2B. *Bottom*: collapsing these measurements based on $\chi$ onto a master curve, for points where $\xi_\eta \geq 0.6$.

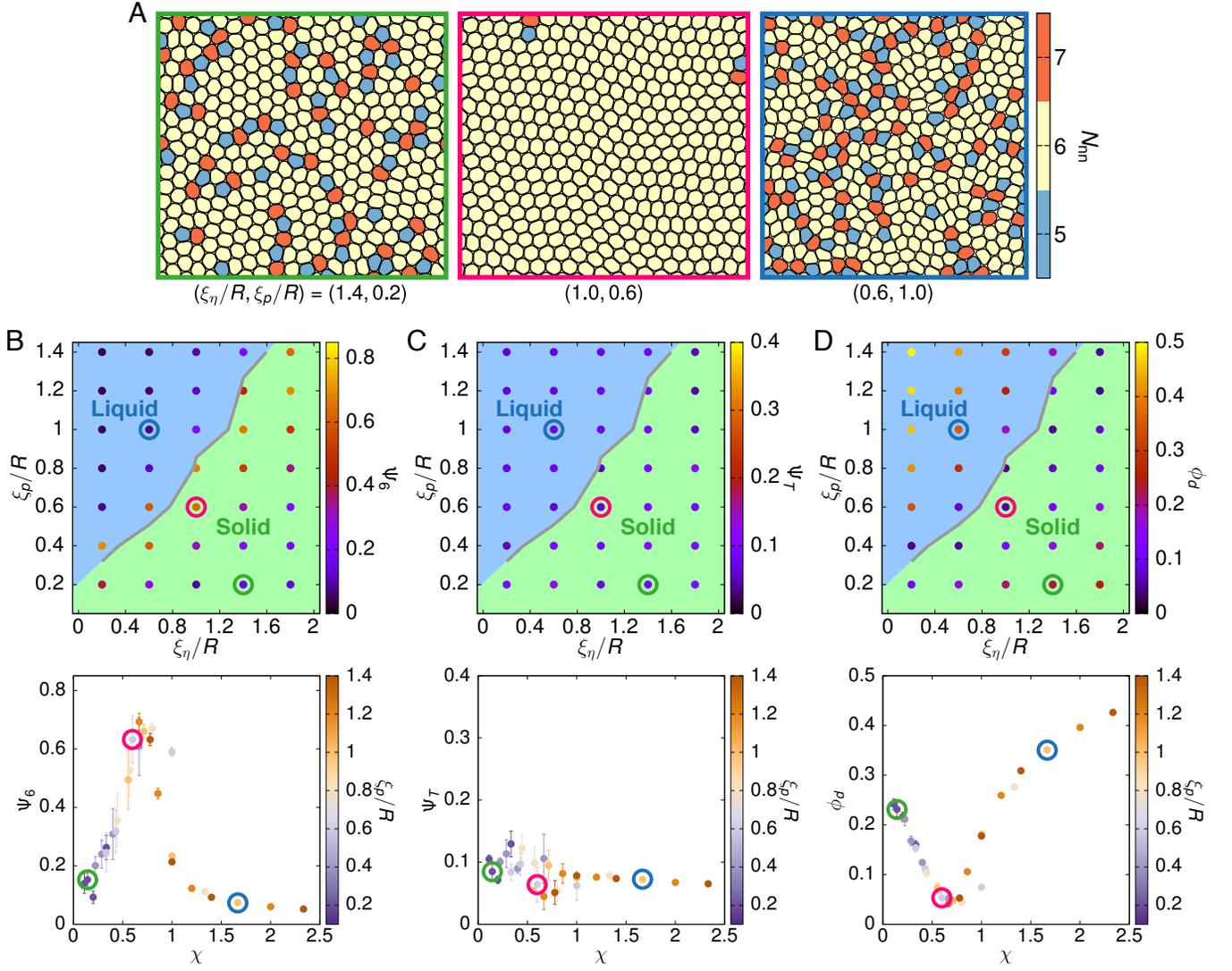

Figure S5. Quantifying the degree of hexatic and translational order for the case of $N = 400$ cells that are initialized randomly. (A) Simulation snapshots of three points in the phase space, with cells colored by their number of nearest neighbors $N_{nn}$ (i.e., 5- and 7-fold disclinations are marked in blue and orange, respectively). (B)–(D) *Top*: Measurements of the (B) hexatic order parameter $\Psi_6$, (C) translational order parameter $\Psi_T$, and (D) fraction of cells with disclinations $\phi_d$ superposed on the solid-liquid phase diagram shown in Fig. S2C. *Bottom*: collapsing these measurements based on $\chi$ onto a master curve, for points where $\xi_\eta \geq 0.6$.

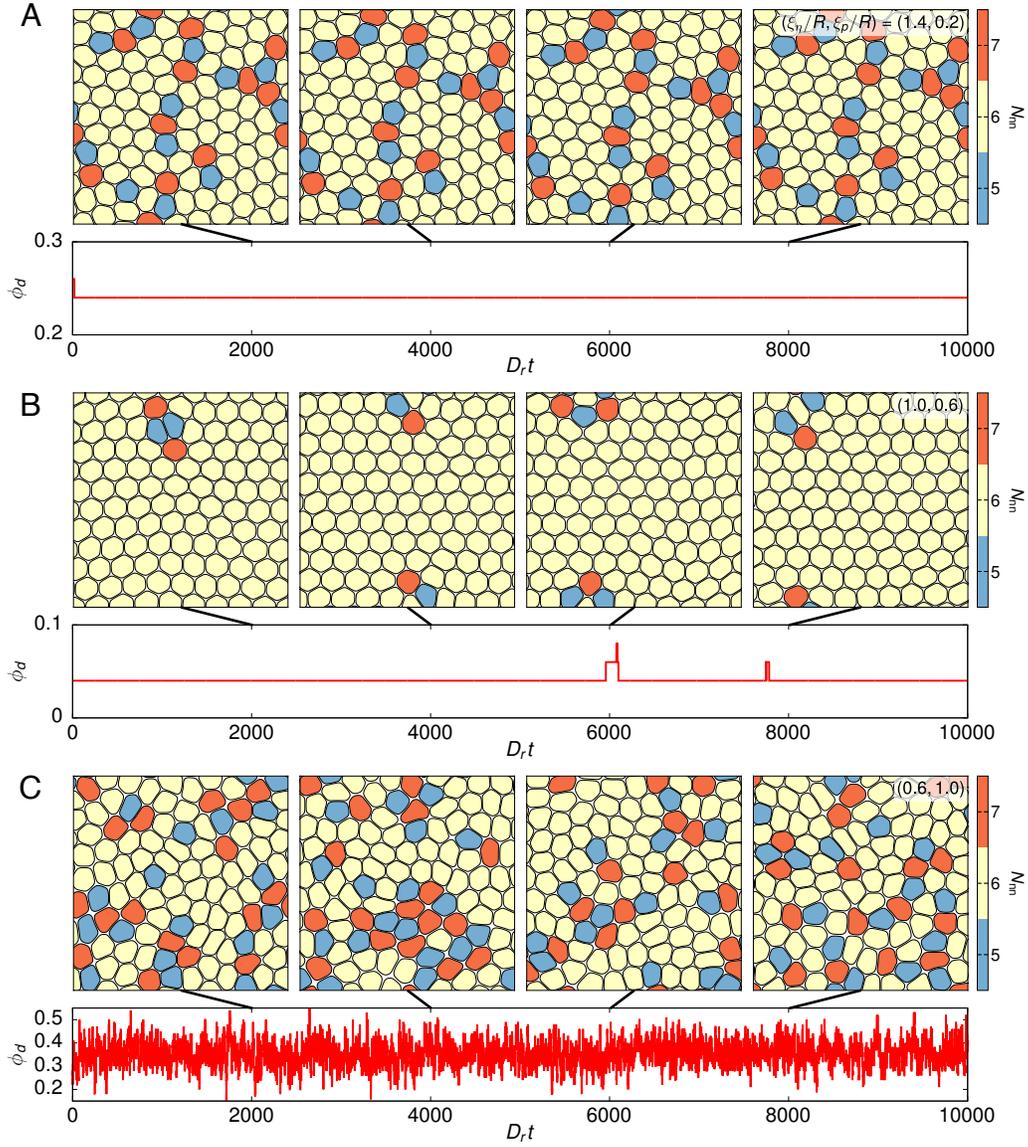

Figure S6. Dynamics of 5-7 disclinations are distinct between liquid and solid phases. (A)–(C) Simulation snapshots showing the number of nearest neighbors $N_{nn}$ for each cell at various time points (*top*) and the time series reporting the fraction of cells with hexatic disclinations $\phi_d$ (*bottom*) for three parameter points, corresponding to (A) the glassy solid regime, (B) the regime with high hexatic order, and (C) the liquid phase. See Fig. S2A for the location of these points within the solid-liquid phase diagram. Data shown here are for the case of $N = 100$ cells that are initialized randomly.

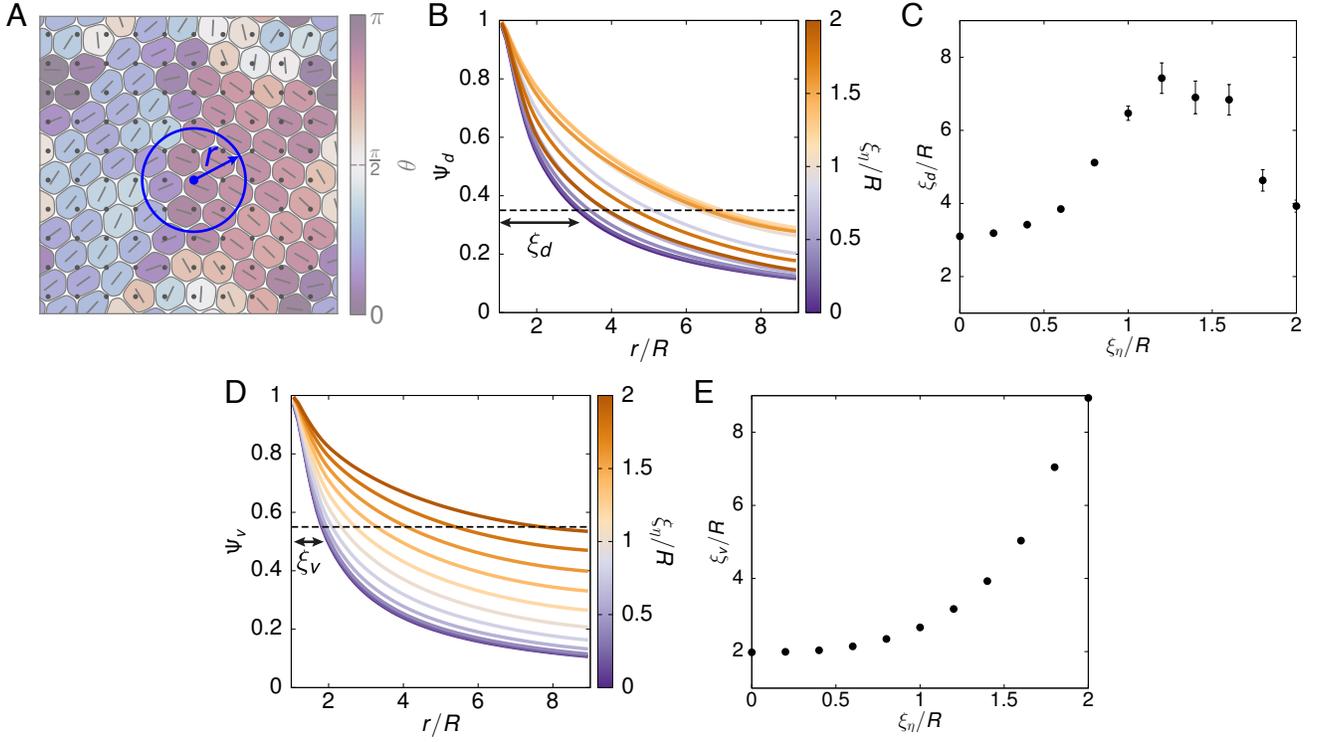

Figure S7. Quantifying the extent of local nematic alignment in cell deformation and the correlation of cellular flow. (A) Schematics explaining how the local nematic orders $\Psi_{d,v}$ are measured and used for extracting a characteristic length. Here, for each grid point (e.g., the one marked in blue), we compute the order parameter for cells within a radius $r$ from the point and average the results over all grid points. (B) The local nematic order in cell deformation $\Psi_d$ as a function of $r$. The curves are for different $\xi_\eta$, with fixed $\xi_p/R = 0.8$, and the dotted line marks the threshold $\Psi_d^* = 0.35$ used for extracting a characteristic length $\xi_d$. (C) The characteristic length $\xi_d$ of the local nematic domain, as estimated from the curves shown in (B). (D) and (E) Similar to (B) and (C), but for the local nematic order in cell advection velocity $\Psi_v$ (i.e., cellular flow). The threshold used for extracting a characteristic length $\xi_v$ is $\Psi_v^* = 0.55$. Data shown here are for the case of $N = 100$ cells that are initialized randomly.

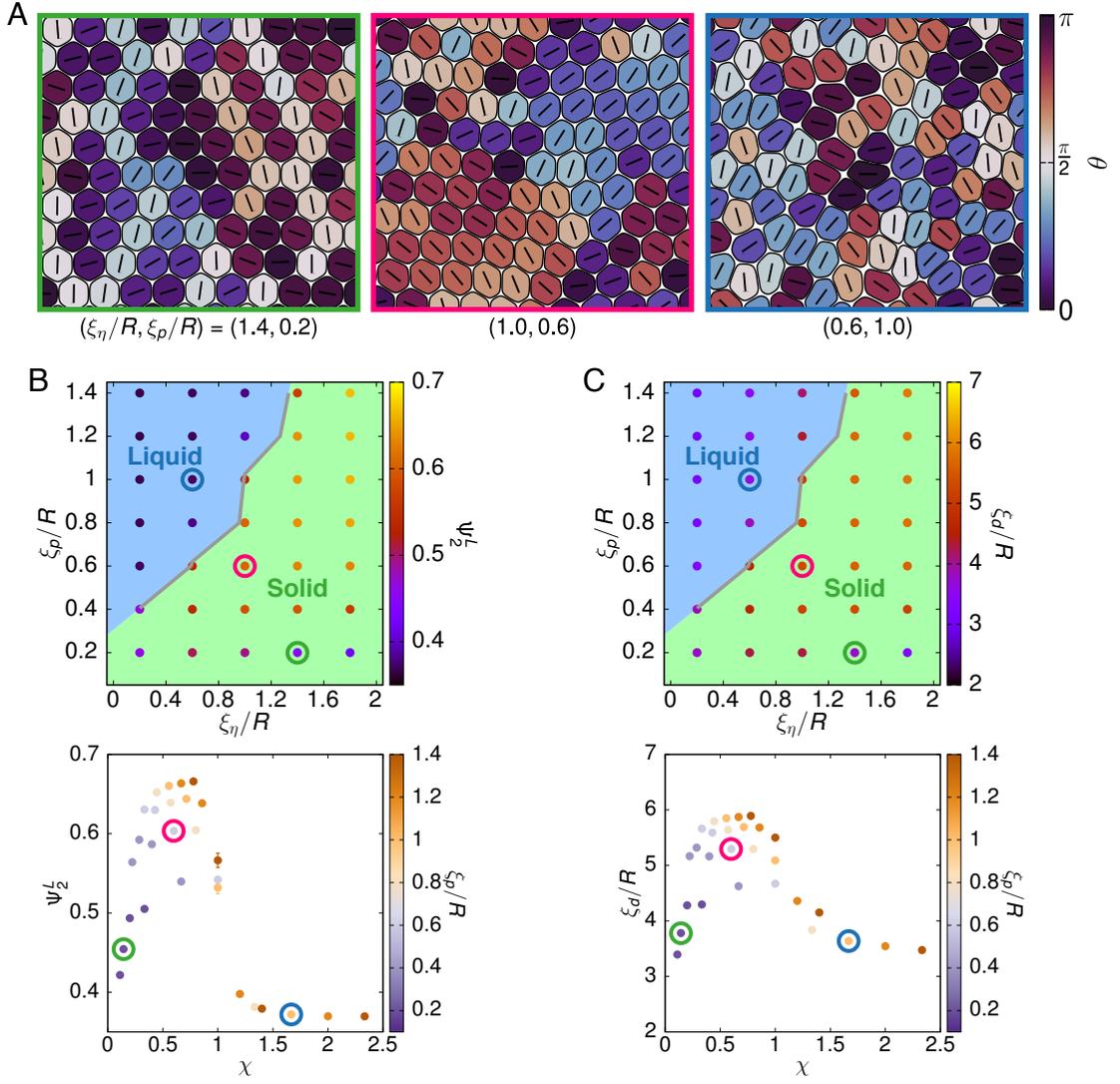

Figure S8. Quantifying the degree of local nematic order in cell deformation for the case of $N = 100$ cells that are initialized on a triangular lattice. (A) Simulation snapshots of three points in the phase space, with cells colored by the angle $\theta$ between their deformation axis (the line within each cell) and the $x$ axis. (B) and (C) *Top*: Measurements of (B) the local nematic order parameter $\Psi_2^L$ and (C) the characteristic length $\xi_d$ of this local nematic order superposed on the solid-liquid phase diagram shown in Fig. S2B. *Bottom*: collapsing these measurements based on $\chi$ onto a master curve, for points where $\xi_\eta \geq 0.6$.

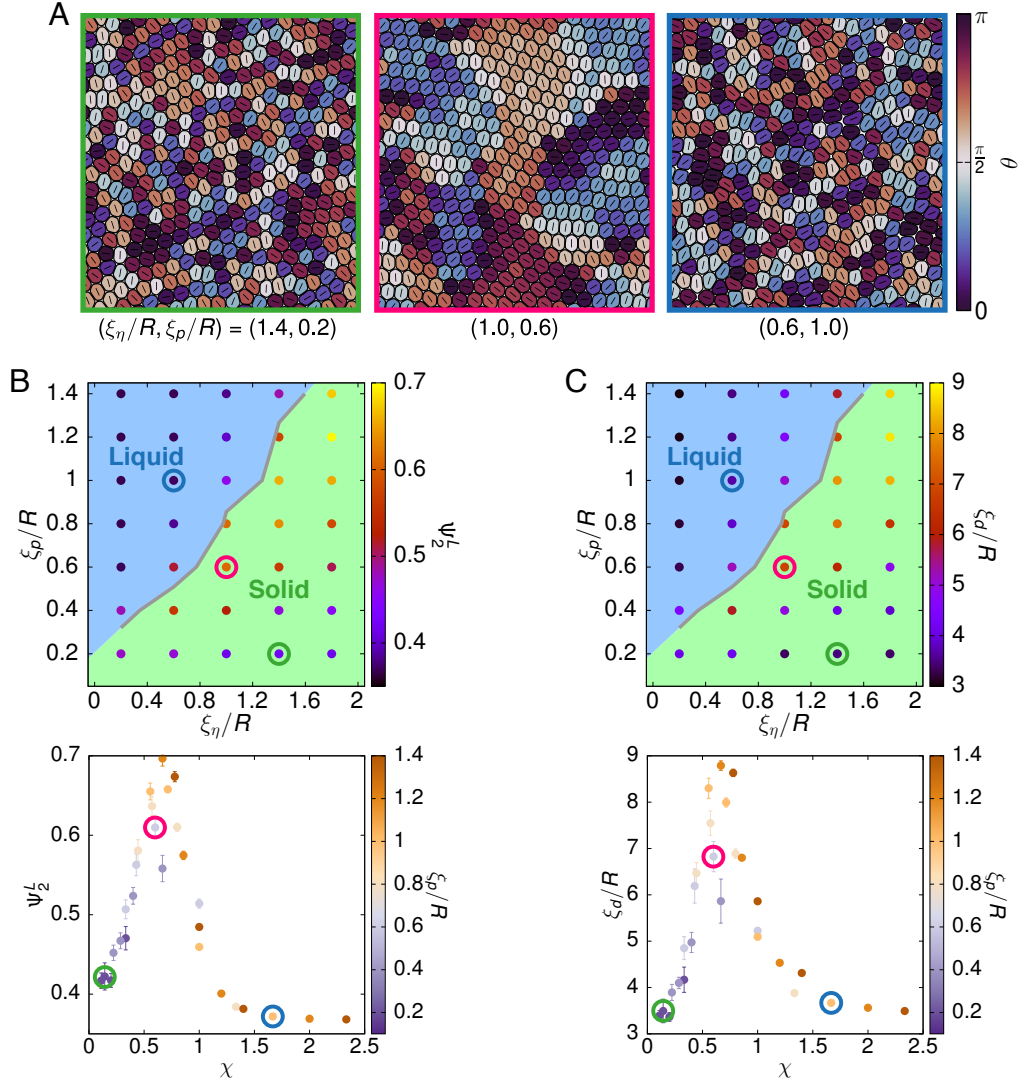

Figure S9. Quantifying the degree of local nematic order in cell deformation for the case of $N = 400$ cells that are initialized randomly. (A) Simulation snapshots of three points in the phase space, with cells colored by the angle $\theta$ between their deformation axis (the line within each cell) and the $x$ axis. (B) and (C) *Top*: Measurements of (B) the local nematic order parameter $\Psi_2^L$ and (C) the characteristic length $\xi_d$ of this local nematic order superposed on the solid-liquid phase diagram shown in Fig. S2C. *Bottom*: collapsing these measurements based on $\chi$ onto a master curve, for points where $\xi_\eta \geq 0.6$.

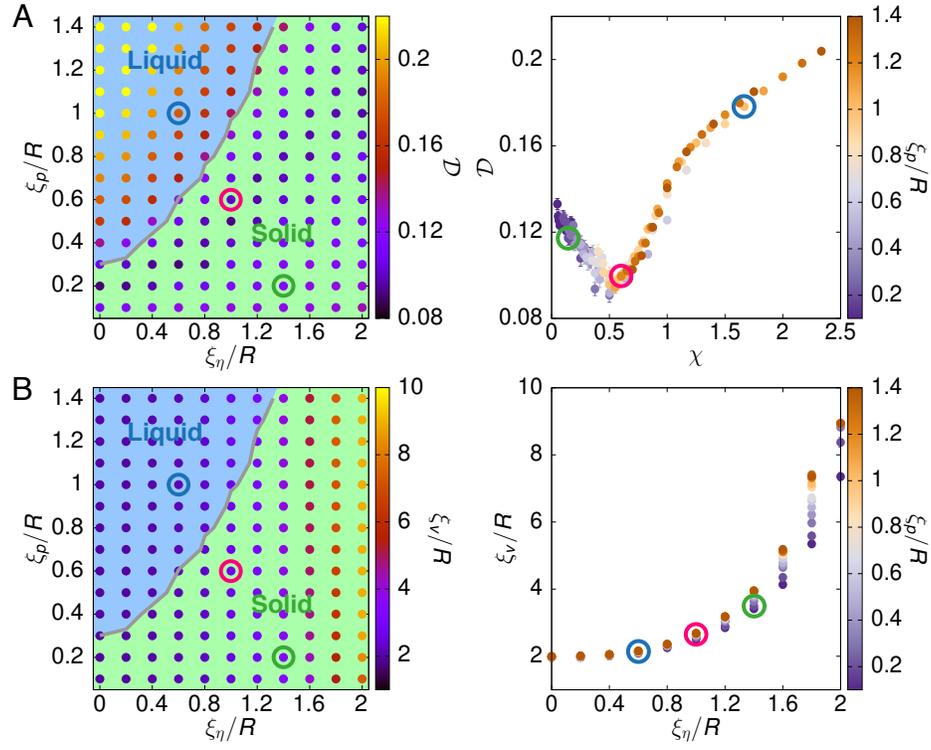

Figure S10. The global degree of cell deformation $\mathcal{D}$ (A) and the characteristic length $\xi_v$ of the local nematic order in cell advection velocity (B; see also Figs. S7D,E) across the phase space. *Left*: measurements of these observables superposed on the solid-liquid phase diagram shown in Fig. 1B. *Right*: collapsing these measurements based on $\chi$ onto a master curve, for points where $\xi_\eta \geq 0.6$. Data shown here are for the case of $N = 100$ cells that are initialized randomly.



\end{document}